\documentclass[preprint2]{aastex}

\shorttitle{UCAC4: A Search for Our Stellar Neighbors}
\shortauthors{Finch}

\begin{document}

\title{UCAC4 Nearby Star Survey: A Search for Our Stellar Neighbors}
\author{Charlie T. Finch, Norbert Zacharias}

\email{finch@usno.navy.mil}

\affil{U.S. Naval Observatory, Washington DC 20392--5420}

\author{John P. Subasavage}
\affil{U.S. Naval Observatory, Flagstaff, AZ 86005--8521}

\author{Todd J. Henry}
\affil{RECONS Institute, Chambersburg, PA 17201}

\author{Adric R. Riedel}
\affil{Hunter College, New York, NY 10065}
\affil{American Museum of Natural History, New York, NY 10024}


\begin{abstract}

We use data from the U.S. Naval Observatory fourth CCD Astrograph
Catalog (UCAC4) in combination with photometry from the AAVSO
Photometric All-Sky Survey (APASS) and Two Micron All-Sky Survey
(2MASS) to identify stars within 25 pc of the Sun.  A sample of nearby
stars with accurate trigonometric parallaxes from the Research
Consortium On Nearby Stars (RECONS) is used to generate a set of 16
new photometric color-$M_{K{_s}}$ relations that provide distance
estimates with uncertainties of 15\%.  This work expands the available
suites of well-calibrated photometric distance relations that can be
used to identify nearby stellar systems.

The distance relations are used with quality cuts to extract an
initial sample of stars from the UCAC4 estimated to be within 25 pc.
Color, proper motion and existing literature sources are then used to
obtain a clean sample of red dwarfs, while limiting the amount of
contamination from background giants, resulting in a sample of 1761
candidate nearby stars within 25 pc.  Of these, 339 are new
discoveries with no previously known published parallax or distance
estimate, primarily with proper motions less than 0.2 arcsec/year.
Five stars are estimated to be within 10 pc, with the nearest, TYC
3980 1081 1 with V$=$ 10.50, estimated to be at 5.93 pc.  That
several hundred new stars have been revealed so close to the Sun
illustrates once again that there is considerable work yet to be done
to map the solar neighborhood, and that additional nearby stars are
likely still to be discovered.

\end{abstract}


\section{INTRODUCTION}

A comprehensive census of the nearby stars is required to determine
accurate stellar luminosity and mass functions in the solar
neighborhood, and is vital to our understanding of the distribution of
stellar mass among different types of stars throughout the Galaxy.  In
addition, because of their proximity to the Sun, the nearest stars are
the most accessible for surveys of stellar activity, ages,
multiplicity, and exoplanets.  A volume-limited sample of stars is
dominated by red dwarfs that make up more than 70\% of stellar objects
in our Galaxy \citep{TSNXVII}.  Still, red dwarfs are currently
under-sampled by volume because of their low luminosities, coupled
with the magnitude limits of current all-sky surveys and the
astrometric limitations of scanned plates.  Ground-based surveys like
the U.S. Naval Observatory Robotic Astrometric Telescope (URAT)
\citep{URAT} currently underway \citep{urataas} will pick up stars
down to 18th magnitude, while space missions like Gaia (launched in
December 2013) will reach to 20th magnitude.

Significant discoveries of nearby stars have been made in the past
decade by searching the Digitized Sky Survey in the northern sky
\citep{Lepine3,Lepine5}, the SuperCOSMOS Sky Survey in the southern
sky \citep{TSNVIII,TSNX,TSNXII,TSNXV,TSNXVIII,TSNXXVII,TSNXXV}, and
\citep{2000A&A...353..958S,2002ApJ...565..539S}, and the USNO CCD
Astrograph Catalog (UCAC), also in the southern sky \citep{upm1,upm2}.
All of these studies rely primarily on searching for stellar systems
with detectable proper motions.  Nonetheless, currently available
datasets have yet to be completely searched for nearby stars,
particularly those with small proper motions.

In this paper we search the more than 100 million sources in the
U.S. Naval Observatory fourth CCD Astrograph Catalog
\citep[UCAC4;][]{UCAC4}, which reaches to 16th magnitude, to reveal
nearby stars.  This all-sky survey takes advantage of the newly
released astrometric results from the UCAC4 along with merged optical
photometry from the American Association of Variable Star Observers
(AAVSO) Photometric All Sky Survey (APASS) and infrared photometry
from the Two Micron All-Sky Survey (2MASS).  With the addition of the
APASS photometry in the UCAC4 release, we are now able to create a
suite of photometric color-M$_{K{_s}}$ relations that provide distance
estimates to nearby red stars with uncertainties of 15\%.  Using this
suite of relations, we have combed the UCAC4 catalog for candidate
nearby stars to reveal new candidates that were missed in previous
searches, most of which required that the stars have detectable proper
motions.  In $\S$2 we describe the datasets and creation of the
photometric color-M$_{K{_s}}$ relations that incorporate $BVgriJHK_s$
photometry and high-quality trigonometric parallaxes.  We outline the
search for nearby star candidates in UCAC4 in $\S$3, describe the
results in $\S$4, and provide concluding remarks in $\S$5.

\section {Data and Techniques}
\subsection {UCAC4 and ALLWISE}

The UCAC4 catalog, released in August 2012, is the fourth and final
version of the UCAC project.  The UCAC4 is an updated version of the
UCAC3 \citep{UCAC3} that brings the positional system closer to that
of UCAC2 \citep{UCAC2} after a final investigation of the magnitude
equation corrections.  UCAC4 also has many bug fixes and utilizes the
Lick Northern Proper Motion (NPM) data in place of the previously used
Schmidt plate data in the north, resulting in reduced systematic
errors in proper motions north of about $-$20$^\circ$ declination.
UCAC4 contains more than 113 million entries, of which nearly 51
million have APASS $BVgri$ photometry and nearly 110 million have
2MASS $JHK_s$. We used a three arcsec match radius in the development
of the UCAC4 catalog for inclusion of the 2MASS and APASS suplimental
photometric data.  Further details can be found in \citep{UCAC4}.
Fewer sources have APASS photometry at present because of the ongoing
progress of the APASS project, which at the time of the UCAC4 release
was currently on Data Release 6 (DR6).  As of the writing of this
manuscript, APASS is at Data Release 7 (DR7).

Included as a supplement for completeness to the photometry we
provide Wide-field Infrared Survey Explorer (ALLWISE) photometry in
the 3.4, 4.6, 12 and 22um mid-infrared bandpasses, hereafter W1, W2,
W3 and W4.

\subsection {Photometric Distances}

To obtain photometric distances for UCAC4 sources, we generated a new
set of 16 photometric color-M$_{K{_s}}$ relations using (a) $BVgri$
optical photometry from APASS, (b) $JHK_s$ near-infrared photometry
from 2MASS, (c) nearby red dwarfs with high-quality trigonometric
parallaxes (defined as having errors of less than 5 milliarcseconds
from the Research Consortium On Nearby Stars (RECONS)\footnote{\it
  www.recons.org} group, and (d) a set of M dwarfs with spectral types
M6.0 V to M9.5 V within 25 pc referred to as the supplemental sample
in \citet[hereafter TSNX]{TSNX}.  The method used is similar to that
in TSNX, in which the M$_{K{_s}}$ band is used with $UBVRIJHK_s$
photometry to obtain 12 relations.  The M$_{K{_s}}$ band is selected
here to boost the number of very red stars having photometry in the
band used for luminosities (e.g. fewer very red stars have $V$
photometry), and to mitigate the problem of interstellar reddening
(although minimal within 25 pc).  The input sample contains 110 stars
from RECONS, a 64 star sample from \citet[hereafter TSNXXII]{TSNXXII},
and 31 stars from the very red supplemental list, bringing the total
sample to 205 stars having accurate trigonometric parallaxes.

We matched the input list of 205 parallax stars with the UCAC4 catalog
to obtain proper motions and $BVgriJHK_s$ photometry.  We filled in
missing photometric data (photometry not found in UCAC4) using the
APASS website, VizieR, or Aladin as needed.  Aladin was also used to
verify stars by eye to eliminate any source misidentifications, of
which none were found.  All parallaxes were updated with the most
recent results from the RECONS database, and APASS photometry was
updated using DR7 web access when available.  All stars in close
multiple systems, known subdwarfs, known white dwarfs, stars without
enough photometric data for at least seven relations, stars with
photometric errors greater than 0.10 magnitude, and stars brighter
than the APASS saturation magnitude limits (defined in Table
\ref{APASSlimit}) were removed from the list, leaving a total of 168
stars for the color-$M_{K{_s}}$ relations.  This list was then given a
final spot check by eye using Aladin and specific catalogs to verify
photometry and parallaxes.

\begin{figure}
  \epsscale{1.00}
  \includegraphics[angle=0,scale=0.5]{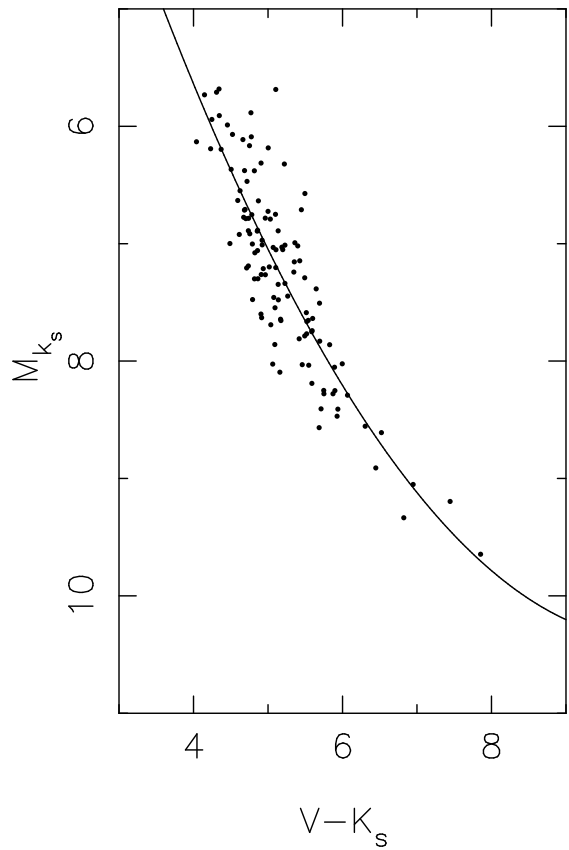}
  \caption{Example color-$M_{K{_s}}$ fit for the ($V-K_s$) color
    relation.}\label{mks}
  \end{figure}

Combinations of eight filter bands provide a total of 28 possible
color-$M_{K{_s}}$ relations.  Of these, 16 were deemed to be reliable,
in that there was at least a one magnitude range in color using the
sample of stars available for a fit, and there was a reasonable
correlation with $M_{K{_s}}$; an example is shown in Figure \ref{mks}
for ($V-K_{s}$).  A second-order fit was used for all relations, with
higher orders not showing any meaningful improvements.  In Table
\ref{crel}, we give the applicable color range, number of stars used,
the fit coefficients and the Root Mean Square (RMS) values for each of
the 16 color relations used in this paper.  Some stars do not have the
complete set of eight-band photometry available or might lie outside
of the applicable color range; thus, the number of stars used for each
fit is less than the sample of 168 stars.  The equations used for the
relations have the following format, shown here for the ($V-K_{s}$)
relation:

\begin{displaymath}
M_{K_{s}} = - 0.1239(V-K_{s})^{2} + 2.523(V-K_{s}) - 2.472 
\end{displaymath}

\noindent These relations are similar to those in TSNX in the number
of stars used ($\sim$100--140) and RMS values ($\sim$0.4 mag).  Both
suites of relations would benefit by increasing the number of very red
stars to extend the applicable ranges of each fit, such as those
reported in \citep{Dieterich}.

\begin{figure}
  \epsscale{1.00}
  \includegraphics[angle=0,scale=0.5]{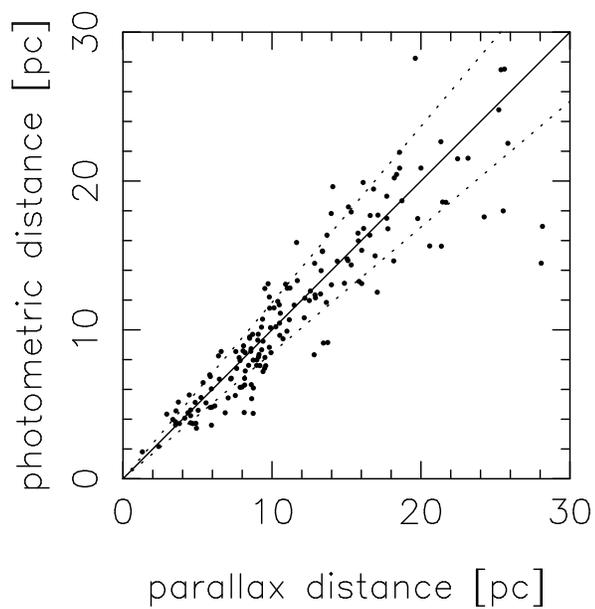}
  \caption{Comparison of the photometric distance estimates and
    trigonometric distances for the 168 stars used in the creation of
    the new suite of relations. The solid line indicates perfect
    agreement, while the dotted lines represent the average 15.5\%
    error. }\label{compare1}
  \end{figure}

The accuracy of the distance estimates has been evaluated by running
the sample of 168 stars used to construct the fits back through the
final relations.  The resulting average distance errors of the 16
estimates, i.e.~the differences between the photometric distance
estimates and the distances from the trigonometric parallaxes, is
15.5\%.  This is virtually identical to the results of TSNX, in which
the average error for the 12 relations was found to be 15.3\%.  We
show a comparison of the photometric distance estimates and the
trigonometric distances in Figure \ref{compare1}.  Of the 168 stars
used to construct the relations, 101 (60.1\%) lie within the adopted
error of 15.5\% and 149 (88.7\%) lie within two times the adopted
error of 15.5\%.  All distance estimate errors reported in this paper
include the 15.5\% external error from the fits and the standard
deviation from the up to 16 distance estimates for a given star, added
in quadrature.

\section {Search for Nearby Stars in UCAC4}

We used a series of four cuts of the UCAC4+APASS+2MASS databases to
extract a candidate list of stars within 25 pc.  In the first cut,
sources were extracted from UCAC4 that met the following criteria:

\begin{itemize}
\item{Target must have photometry in at least two filters in the APASS
  catalog, with photometry errors (apase) $\le$ 0.10 mag.}
\item{Target must have photometry in all three $JHK_s$ filters in the
  2MASS catalog, with photometry errors (e2mpho) $\le$ 0.10 mag.} 
\item{Target must not have a UCAC object flag (objt) $=$ 1 or 2,
  indicating it is near an overexposed star or a streaked object.}
\item{Target must have a valid, non-zero proper motion in UCAC4.}
\item{Target must have a Lyon-Meudon Extragalactic DAtabase (LEDA)
  galaxy flag (leda) $=$ 0 and a 2MASS extended source flag (2mx) $=$
  0 --- both indicate a point source.}
\end{itemize}

There were 25,865,591 sources output from this query.  Those were then
run through the suite of photometric distance relations and a second
cut was performed that extracted only those for which (a) at least
seven of the 16 relations could be applied, and (b) distances were
estimated to be within 25 pc, yielding a list of 381,054 candidate
nearby stars.

To build a dataset for further investigation, this list was then
cross-matched using a 30 arcsecond radius with the RECONS 25 Parsec
Database, the Hipparcos catalog via VizieR, SIMBAD, and selected
journal papers to gather names, spectral types, parallaxes, distance
estimates, and other useful information about these stars.  During
this process, we noticed that many candidates were either associated
with or within 30 arcseconds of x-ray sources; these candidates are
noted as such in the Tables.  A list of journal publications and
online catalogs used during this search to help build the dataset is
given in Table \ref{search}.

Removing background giants from this large candidate nearby star
sample was accomplished using two more cuts.  The third cut identifies
giants using $J-K_s$ and $V-K_s$ colors, similar to the techniques
outlined in \citet{Adric} for candidates extracted using SuperCOSMOS
and 2MASS data.  The fourth cut uses the UCAC4 proper motions in a
Reduced Proper Motion diagram (RPM), as seen in \citet{Lepine1} and
similar to the RPM diagrams in
\citet{TSNXII,TSNXV,TSNXVIII,TSNXXVII,TSNXXV} to further eliminate any
contamination.

\begin{figure}
  \epsscale{1.00}
  \includegraphics[angle=-90,scale=0.5]{fig3.ps}
  \includegraphics[angle=0,scale=0.5]{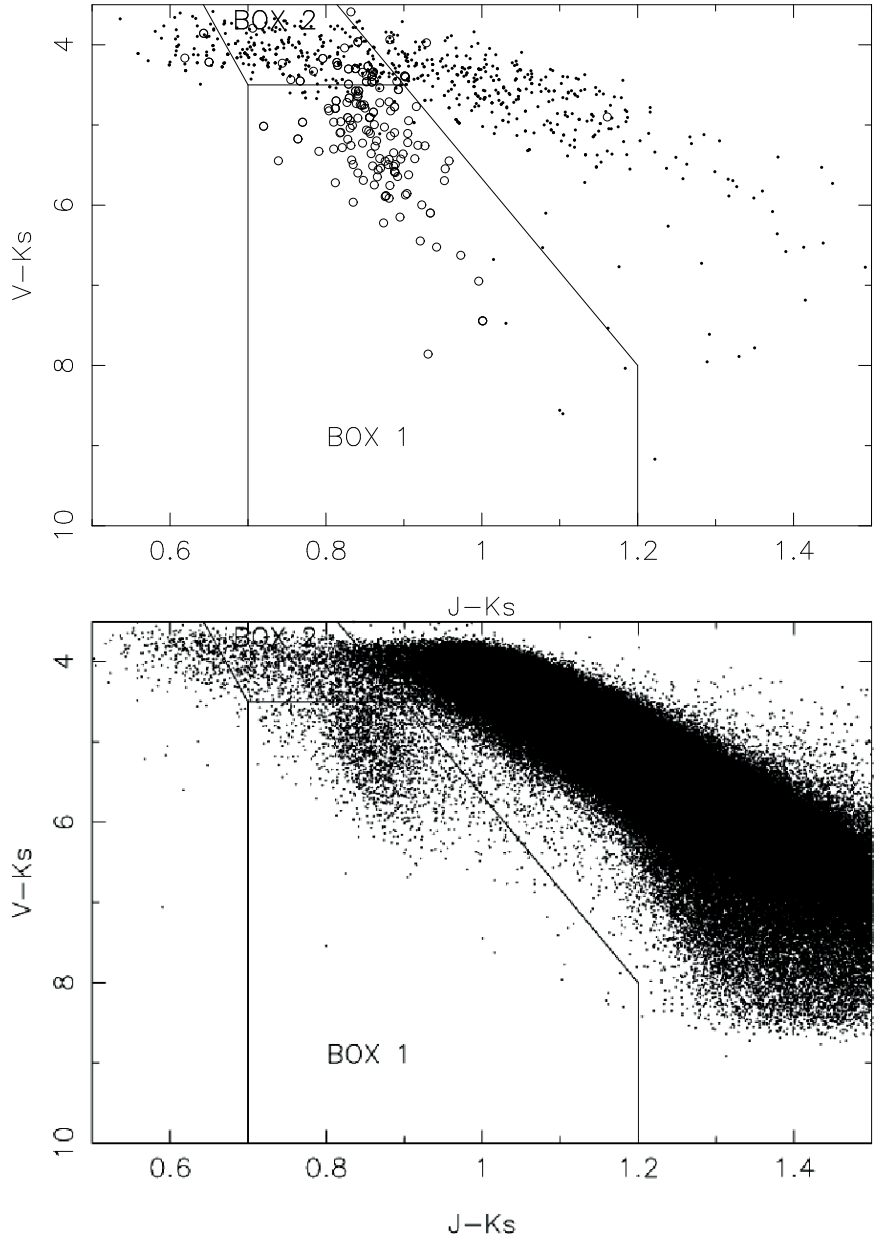}
  \caption{$V-K_s$ vs.~$J-K_s$ color-color plots showing (top) a
    sample of known giants (dots) and M dwarfs (open circles) from
    the 381,054 candidate nearby star sample, including a supplement
    sample from RECONS to help construct BOX 1 and BOX 2, and
    (bottom) the entire 381,054 stars from the candidate nearby star
    sample.}\label{color}
  \end{figure}

In the top panel of Figure \ref{color} we show a sample of known
giants (dots) and M dwarfs (open circles) pulled from the 381,054
candidate nearby star list, supplemented with a known sample of dwarfs
from RECONS.  Using $V-K_s$ and $J-K_s$ colors, we constructed two
boxes similar to those in \citet{Adric} that likely contain M dwarfs
(vertices are given in Table \ref{boxes}).  Box 1 contains the most
likely M dwarf candidates, while box 2 is more likely to be
contaminated by giants.  The bottom panel in Figure \ref{color} shows
the entire 381,054 sample of nearby star candidates, and after the
third cut, we reduce the sample to 4424 candidates that are found in
the two boxes.

To further eliminate background giants, the fourth cut utilizes proper
motions from UCAC4, combined with photometry from APASS and 2MASS, to
plot an RPM diagram.  We used a modified distance modulus equation,
where $\mu$ in arcsec/year is substituted for distance:

\begin{displaymath}
H_v = V+5\log\mu+5.
\end{displaymath}

\begin{figure}
  \epsscale{1.00}
  \includegraphics[angle=-90,scale=0.5]{fig5.ps}
  \includegraphics[angle=0,scale=0.5]{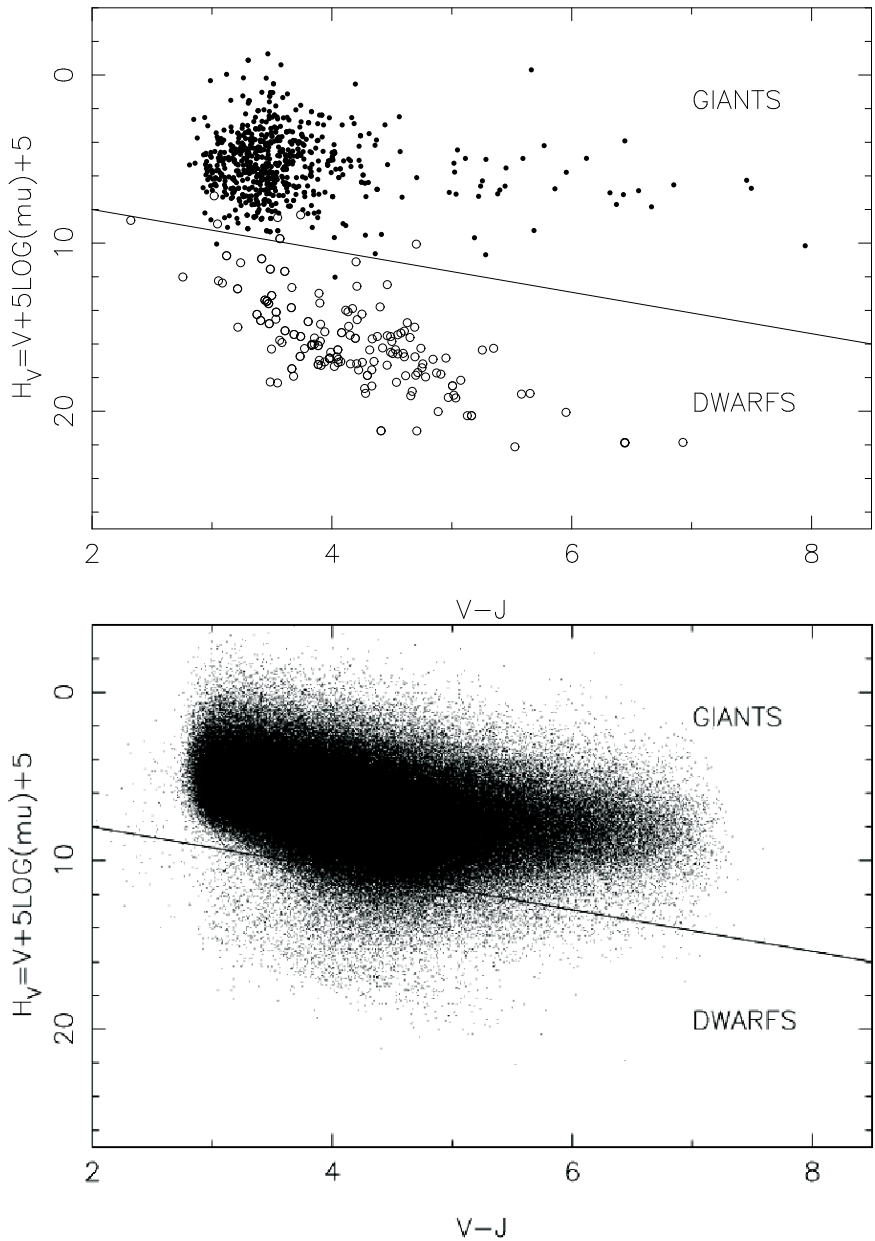}
  \caption{Reduced Proper Motion diagram showing (top) a sample of
    known giants (dots) and M dwarfs (open circles) from the 381,054
    candidate nearby star sample, supplemented with a sample of stars
    from RECONS having known parallaxes to help construct a
    delimiting line, and (bottom) the 381,054 stars from the
    candidate nearby star sample.}\label{hr}
  \end{figure}

An RPM diagram is useful in separating dwarfs and giants because
distant giants tend to have very slow proper motions compared to
nearby dwarfs.  In the top plot of Figure \ref{hr}, we show an RPM
diagram of the same known sample of giants (dots) and M dwarfs (open
circles) used in Figure \ref{color}.  A delimiting line is drawn to
separate the two classes of stars, with vertices at (2,8) and
(8.5,16).  In the bottom plot of Figure \ref{hr}, we show the entire
initial sample of 381,054 candidates, and use the RPM cut to eliminate
another 2408 presumed giants from the 4424 candidates in the two boxes
outlined in Figure \ref{color} for the color cut, yielding 2016
candidate nearby stars.  After removing 255 duplicate entries which we
found to be the only duplicates in the entire initial 381,054 sample
we are left with a manageable list of 1761 candidate nearby stars.

After the color and RPM cuts, all stars having known parallaxes (669),
distance estimates in a published paper (749), or known to be giants
(4) were extracted, leaving 339 new candidate nearby stars within 25
pc.  Details for all 1761 nearby candidates are given in Tables
\ref{listnew} (339 new discoveries) and \ref{recover} (1422 recovered
stars), where we give the system names, RA and DEC coordinates, proper
motions in RA and DEC with associated errors, $BVgriJHK_sW1W2W3W4$
photometry, distance estimates with associated errors, the number of
color relations used for the estimates (typically all 16) and relevant
notes.  In Table \ref{recover}, we provide additional columns for
published distances and the type of distance (P = photometric
distance; T = trigonometric distance).  The first 28 lines of Table
\ref{listnew} are provided here sorted by distance and representing
all new candidate nearby star systems out to 15 pc.  The first 15
lines of Table \ref{recover} are provided here sorted by RA.  The full
versions of Tables \ref{listnew} and \ref{recover} can be found in the
electronic version of this paper.  All positions, proper motions and
$BVgriJHK_s$ come directly from the UCAC4 with some APASS photometry
updated to the recent DR7 release for those stars with missing
photometry (current release data can be found at {\it
  www.aavso.org/apass}). If no photometry was found or the photometry
between 2MASS and APASS was not in agreement in the UCAC4 then the
APASS and 2MASS databases were searched manually using either VizeiR
or the APASS websites to check for consistency and to fill in gaps.

As a final check, the 339 new nearby candidates were evaluated by eye
using Aladin to reveal false detections and to verify proper motions.
Of these, 101 did not show any detectable proper motion (noted in
Table \ref{listnew}).  Further investigation showed that all but six
of the 101 objects had fewer than one arcsecond of total UCAC4 proper
motion given the epoch spread of the digitized plates used for visual
inspection.

Finally, it is evident from looking at Figures \ref{color} and
\ref{hr} that a few {\it bona fide} M dwarfs are seen outside the
boxes or above the delimiting line.  As a result of the cuts, eight
known M dwarfs were omitted from a sample of 181 known M dwarfs, the
same sample shown in the top panels of Figures \ref{color} and
\ref{hr}, which indicates that we have omitted $\sim$4\% of nearby
stars by our search method.  Ultimately, given that our goal is a
clean sample of dwarfs, we have omitted a few in favor of minimizing
contamination by giants.

\section {Results}

The 339 new nearby star candidates are all estimated to be within 25
pc of the Sun, including five predicted to be within 10 pc.  Of the
new 339 nearby star candidates, 218 are previously unknown proper
motion stars and have been given a USNO Proper Motion (UPM) name
designation for this paper, consistent with previous discoveries from
our UCAC efforts \citep{upm1,upm2}.

\begin{figure}
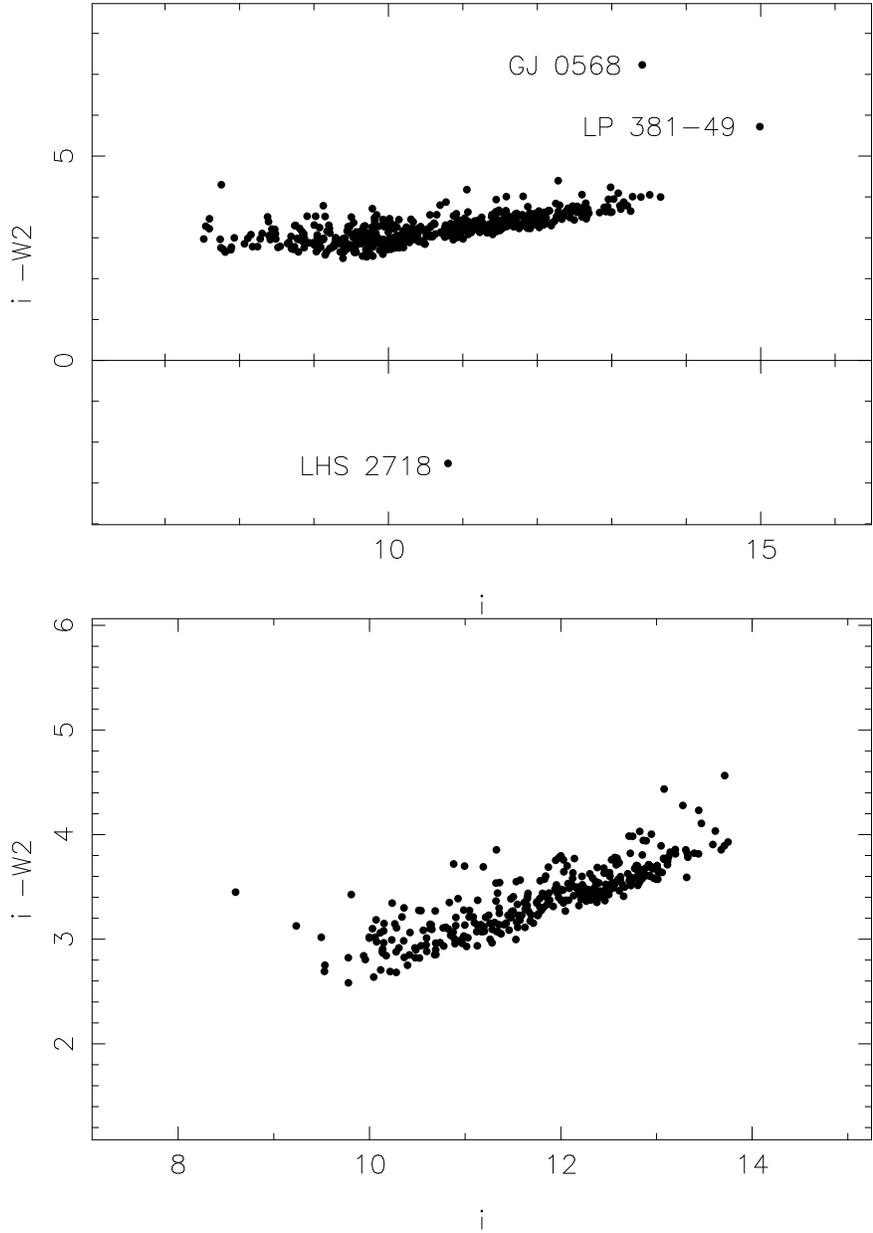

  \epsscale{1.00}
  \includegraphics[angle=-90,scale=0.5]{fig7-1.ps}
  \includegraphics[angle=-90,scale=0.5]{fig8-1.ps}
  \caption{i-W2 vs. i color-magnitude plot of 479 known stars from
    Table \ref{recover} having ALLWISE photometry and a trigonometric
    parallax (top) and the same color-magnitude plot showing the 339
    new nearby star candidates from Table \ref{listnew} (bottom) for
    comparison.}\label{wise1}
  \end{figure}

In Figure \ref{wise1}, we show a color-magnitude plot of 479 known
stars with trigonometric parallaxes in Table \ref{recover} having
available ALLWISE photometry (top) and the same color-magnitude plot
of the 339 new nearby star candidates from Table \ref{listnew}
(bottom) for comparison. In Figure \ref{wise2}, we show a color-color
plot of 479 known stars with trigonometric parallaxes in Table
\ref{recover} having available ALLWISE photometry (top) and the same
color-color plot of the 339 new nearby star candidates from Table
\ref{listnew} (bottom) for comparison. The known sample (top) in
Figures \ref{wise1} and \ref{wise2} show some outliers from the bulk
of the sample which have been labeled on the plot. The new nearby star
candidates reported in this paper compare well with the bulk of the
known sample.

In Figure \ref{sky}, we show the sky distribution of the 339 new
nearby star candidates.  This plot shows that during this all-sky
survey, more new nearby star candidates were discovered in the
southern sky than in the north.  This is because the southern
hemisphere has historically been less rigorously searched for stars
with proper motions less than $\sim$0.18$\arcsec$/yr, the cutoff of
Luyten's all-sky survey \citep{1980PMMin..55....1L}.  In
Figure~\ref{hist1}, we show a proper motion histogram of the newly
discovered nearby star candidates in 0$\farcs$02 yr$^{-1}$ bins,
highlighting with dark bars those having distance estimates within 15
pc.  This illustrates that most of the new nearby stars we have
revealed are found in the slower proper motion regimes, yet each is
estimated to be closer than 25 pc.

We used the RECONS 25 Parsec Database to construct a catalog of 320
stars with accurate trigonometric parallaxes placing them within 10
pc.  Many of these objects are very bright (AFGK stars) or very faint
(brown dwarfs), and are therefore outside the limits of our current
search, which targets red dwarfs.  A detailed analysis indicates
that 75 did not match up to any UCAC4 entry, 34 did not contain any
APASS photometry in UCAC4, 54 did not have any 2MASS photometry listed
in UCAC4, 75 did not contain photometry with errors less than or equal
to 0.10 mag, 10 did not meet the color range for the relations, nine
did not meet the criteria of using at least seven relations for the
distance estimate, and one was rejected because of an object flag.  Of
the remaining 62 stars that were recovered successfully, 51 (82\%)
are estimated to be within 10 pc using the relations presented here,
and only one is estimated to be beyond 15 pc.  We conclude that
our search is successful in revealing nearby red dwarfs, and that our
new relations provide reasonable distance estimates.

\subsection{Local Statistics}

As of 01 January 2014, there are 270 systems known to be within 10 pc
of the Sun \citep{TSNXVII} and updates on ({\it www.recons.org}).  This
census is extracted from the RECONS 25 Parsec Database, which includes
all systems having trigonometric parallaxes of at least 40 mas and
errors of 10 mas or less, reported in the refereed literature.  Our
survey for nearby stars in the UCAC4 has turned up 13 stars predicted
to be within 10 pc that have no published trigonometric parallaxes.
Eight of the 13 stars have previously published distance estimates,
and here we provide the first distance estimates for the remaining
five candidates: TYC 3980-1081-1 (5.93 pc), L 173-19 (8.47 pc), 2MASS
J20490993-4012062 (8.66 pc), BPS CS 22898-0066 (9.58 pc), and TYC
3251-1875-1 (9.66 pc).

Seven of the 13 stars are already on the RECONS astrometry program
being carried out at the CTIO/SMARTS 0.9m \citep{2005AJ....129.1954J}.
Two have trigonometric parallaxes recently published in
\citep{2014AJ....147...85R} --- 2MASS J07491271-7642065 (L034-026) at
10.6 pc and L 449-001AB at 11.9 pc.  Results for the other five
indicate that four are indeed within 10 pc (and will soon be
published), while the fifth is a young star at 43.2 pc.  The remaining
six stars are either too far north for the astrometry program in Chile
or may be added to determine parallaxes.

Expansion of this core sample of 10 pc stars to those within 25 pc
allows us to determine more robust luminosity and mass functions in
the solar neighborhood, which provide benchmarks for our understanding
of the stellar content of our Galaxy, as well as other galaxies.
Although the color and proper motion cuts described above were
successful at eliminating many background giants, the remaining sample
will still suffer from some contamination.  We evaluate the fraction
of the 339 new nearby candidates likely to be within 25 pc by
examining the 672 having known parallaxes included in the recovered
sample Table \ref{recover}.  Of those, 562 have parallaxes placing
  them within 25 pc, whereas 110 are beyond 25 pc, for a success rate
  of 84\%.  This implies that $\sim$285 of the discoveries are likely
  to be within 25 pc.

As of 01 January 2014, there are 2168 systems known within 25 pc that
have accurate trigonometric parallaxes.  The anticipated 285 new
systems within 25 pc found in this UCAC4 survey may therefore increase
the sample by as much as 13\%.  Those stars that prove to be further
than 25 pc will be overluminous, typically because they are unresolved
multiples and/or young stars.  Thus, while not every one of the new
discoveries first reported here will be a star within 25 pc, even
those beyond 25 pc are likely to be astrophysically compelling
targets.

\section {Conclusions}

We have identified 339 new nearby star candidates within 25 pc via this
search, of which 218 were previously unidentified in {\it any} search
or catalog.  Most of these new discoveries are nearby stars with
proper motions less than 0.2 arcsec/yr, indicating that many of the
Sun's nearest neighbors have been missed to date because of the
emphasis on larger proper motions.

The 16 relations reported here can be used to estimate distances to
nearby red dwarfs within the applicable color ranges given in Table 2.
The relations are quite reliable, as evidenced by the number of nearby
stars recovered during the search.  The primary limitations of the
method are a lack of APASS sky coverage, and that the relations focus
on red dwarfs only.  This search used the UCAC4 catalog as a starting
point, but a new search could be done once the APASS and upcoming URAT
catalogs are complete, providing improved photometric sky coverage and
depth.

Followup trigonometric parallax measurements are required to confirm
which of the nearby star candidates are, in fact, closer than 25 pc.
Given continued observations by the RECONS team from the ground, and
{\it Gaia's} anticipated astrometric accuracy and limiting magnitude,
most or all of these targets should have accurate parallaxes in the
coming years.


\acknowledgments

We thank the entire UCAC team for making this nearby star search
possible.  Special thanks go to members of the RECONS team at for
their support and use of the RECONS 25 pc database.  This work has
made use of the SIMBAD, VizieR, and Aladin databases operated at the
CDS in Strasbourg, France.  We have also made use of data from 2MASS,
APASS, ALLWISE and the ADS service.


\clearpage


 \clearpage
 
 \clearpage
 \clearpage
 
 \clearpage
 
 \clearpage

  \begin{figure}
  \epsscale{1.00}
  \includegraphics[angle=-90,scale=0.5]{fig9.ps}
  \includegraphics[angle=-90,scale=0.5]{fig10.ps}
  \caption{V-i vs. i-W2 color-color plot of 479 known stars from
    Table \ref{recover} having ALLWISE photometry and a trigonometric
    parallax (top) and the same color-color plot showing the 339
    new nearby star candidates from Table \ref{listnew} (bottom) for
    comparison. }\label{wise2}
  \end{figure}
 \clearpage

 \begin{figure}
 \epsscale{1.00}
 \includegraphics[angle=-90,scale=0.40]{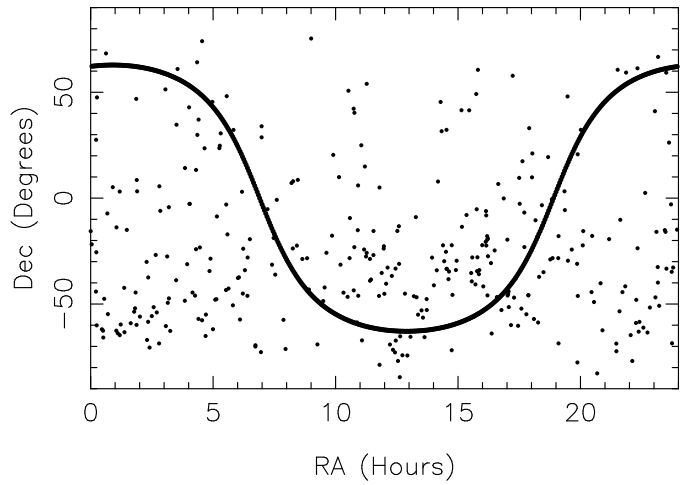}
 \caption{Sky distribution of the 339 new nearby star candidates
   reported in this paper.  The curve represents the Galactic
   plane. }\label{sky}
 
 \end{figure}
 \clearpage
 
 \begin{figure}
 \epsscale{1.00}  
 \includegraphics[angle=-90,scale=0.40]{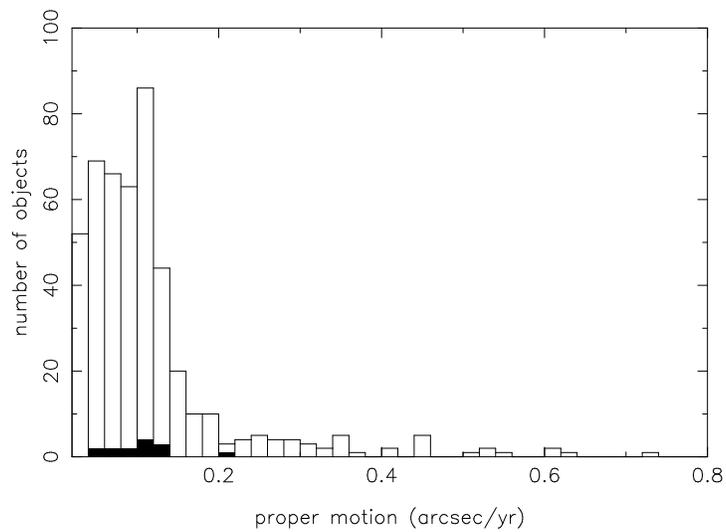}
 \caption{Histogram showing the number of proper motion objects in
   0$\farcs$02 yr$^{-1}$ bins for all new nearby star candidates
   (light shade) and the number of those objects having distance
   estimates within 15 pc (dark shade)}\label{hist1}
 \end{figure}
 
 %
 
 \clearpage



\begin{deluxetable}{cc}
\tabletypesize{}

\tablecaption{APASS photometric saturation magnitude
  limits\tablenotemark{1}\label{APASSlimit}}

\tablewidth{0pt}
\tablehead{
\colhead{Bandpass}             &
\colhead{Saturation Limit}     }

\startdata
  B & 10.25   \\
  V & 10.25   \\
  g & 11.00   \\
  r & 10.00   \\
  i &  9.00   \\

\enddata

\tablenotetext{1} {as of Data Release Six (DR6)}

\end{deluxetable}

\clearpage
\begin{deluxetable}{lccrrrr}
\tabletypesize{}

\tablecaption{Details of the 16 Photometric Distance Relations\label{crel}}

\tablewidth{0pt}
\tablehead{
\colhead{     }                &
\colhead{Color Range     }     &
\colhead{Stars Used}           &
\colhead{Coeff. 1}             &
\colhead{Coeff. 2}             &
\colhead{Coeff. 3}             &
\colhead{rms}                 \\

\colhead{Color}                &
\colhead{[mag]}                &
\colhead{[number]}             &
\colhead{[$\times$ color\emph{$^2$}]} &
\colhead{[$\times$ color]}     &
\colhead{[constant]}           &
\colhead{[mag]}                }

\startdata
V--i  & 1.4 --  3.9  & 113  & $-$0.33800 & $+$3.663 & $+$0.9571 & 0.38   \\
V--J  & 3.1 --  7.0  & 118  & $-$0.14260 & $+$2.551 & $-$1.0870 & 0.40   \\
V--H  & 3.5 --  7.7  & 118  & $-$0.13910 & $+$2.657 & $-$2.3950 & 0.41   \\
V--K  & 4.0 --  8.0  & 118  & $-$0.12390 & $+$2.523 & $-$2.4720 & 0.42   \\
B--i  & 2.8 --  6.0  & 102  & $-$0.13570 & $+$3.091 & $-$1.6910 & 0.38   \\
B--J  & 4.2 --  9.0  & 140  & $-$0.08928 & $+$2.188 & $-$2.6330 & 0.37   \\
B--H  & 4.9 -- 10.0  & 141  & $-$0.09372 & $+$2.347 & $-$4.0930 & 0.39   \\
B--K  & 5.0 -- 10.0  & 141  & $-$0.08031 & $+$2.174 & $-$3.8580 & 0.39   \\
g--i  & 2.2 --  4.5  & 105  & $-$0.13340 & $+$3.619 & $-$1.0110 & 0.39   \\
g--J  & 3.9 --  7.8  & 113  & $-$0.13290 & $+$2.693 & $-$2.8800 & 0.39   \\
g--H  & 4.2 --  8.4  & 113  & $-$0.11760 & $+$2.796 & $-$4.2560 & 0.41   \\
g--K  & 4.5 --  8.8  & 108  & $-$0.13480 & $+$2.637 & $-$4.2120 & 0.41   \\
r--i  & 1.0 --  3.0  & 105  & $-$0.26210 & $+$3.237 & $+$2.9860 & 0.41   \\
r--J  & 2.9 --  6.2  & 107  & $-$0.11530 & $+$2.383 & $+$0.2988 & 0.41   \\
r--H  & 3.4 --  6.8  & 107  & $-$0.28630 & $+$2.502 & $-$0.9758 & 0.42   \\
r--K  & 3.5 --  7.1  & 102  & $-$0.19210 & $+$2.353 & $-$1.0590 & 0.42   \\

\enddata
\end{deluxetable}

\clearpage
\begin{deluxetable}{lr}
\tabletypesize{}

\tablecaption{List of journal publications and online catalogs used
  while searching for information on stars in the candidate nearby
  star list\label{search}}

\tablewidth{0pt}
\tablehead{
\colhead{description}             &
\colhead{Reference}         }

\startdata
 A Catalog of Northern Stars with Annual Proper Motions Larger than 0.15"       & \citep{Lepine0}     \\
 An All-sky Catalog of Bright M Dwarfs                                          & \citep{Lepine1}     \\ 
 Carbon-enhanced metal-poor stars                                               & \citep{Masseron}    \\
 Carbon-rich giants in the HR diagram                                           & \citep{Bergeat}     \\
 Carbon stars from the Hamburg/ESO survey                                       & \citep{Christlieb}  \\
 Characterization of M, L, and T Dwarfs in the Sloan Digital Sky Survey         & \citep{Hawley2}     \\
 Contributions to the Nearby Stars (NStars) Project: The Northern Sample. I.    & \citep{Gray1}       \\
 Contributions to the Nearby Stars (NStars) Project: The Southern Sample        & \citep{Gray2}       \\ 
 Galactic mass-losing AGB stars probed with the IRTS. I.                        & \citep{Bertre1}     \\
 Galactic mass-losing AGB stars probed with the IRTS. II.                       & \citep{Bertre2}     \\
 Hiding in Plain Sight                                                          & \citep{Adric}       \\
 Identification of New M Dwarfs in the Solar Neighborhood                       & \citep{Riaz1}       \\
 Luminous Carbon stars in Galactic Plane                                        & \citep{Kastner}     \\
 Meeting the Cool Neighbors. VII.                                               & \citep{Reid1}       \\
 Meeting the Cool Neighbors. VIII.                                              & \citep{Reid2}       \\
 Nearby Stars from the LSPM-North Proper-Motion Catalog. I.                     & \citep{Lepine3}     \\
 Near-infrared photometry of carbon stars                                       & \citep{Whitelock}   \\
 New High Proper Motion Stars from the Digitized Sky Survey. III.               & \citep{Lepine2}     \\
 Palomar/MSU nearby star spectroscopic survey                                   & \citep{Hawley1}     \\
 Southern infrared proper motion survey. I.                                     & \citep{Deacon1}     \\
 Southern infrared proper motion survey. II.                                    & \citep{Deacon2}     \\
 Spectroscopic Survey of M Dwarfs within 100 Parsecs of the Sun                 & \citep{Bochanski}   \\
 The HIPPARCOS and Tycho catalogues.                                            & \citep{Perryman}    \\
 The Palomar/MSU Nearby Star Spectroscopic Survey. III.                         & \citep{Gizis}       \\
 The Solar Neighborhood. VIII.                                                  & \citep{TSNVIII}     \\
 The Solar Neighborhood. X.                                                     & \citep{TSNX}        \\   
 The Solar Neighborhood. XII.                                                   & \citep{TSNXII}      \\
 The Solar Neighborhood. XV.                                                    & \citep{TSNXV}       \\
 The Solar Neighborhood. XVIII.                                                 & \citep{TSNXVIII}    \\
 The Solar Neighborhood. XXV.                                                   & \citep{TSNXXV}      \\
 The Solar Neighborhood. XXVII.                                                 & \citep{TSNXXVII}    \\
 The Tycho-2 catalogue                                                          & \citep{Hoeg}         \\
 Trigonometric Parallaxes for 1507 Nearby Mid-to Late M-dwarfs                   & \citep{Dittmann}    \\
\enddata
\end{deluxetable}

\clearpage
\begin{deluxetable}{ccc}
\tabletypesize{}

\tablecaption{Verticies of box1 and box2 chosen from Figure
  3\label{boxes}}

\tablewidth{0pt}
\tablehead{
\colhead{}                   &
\colhead{J$-$K\emph{$_s$}}     &
\colhead{V$-$K\emph{$_s$}}     }

\startdata
box 1 & 0.70 &  4.5  \\
      & 0.90 &  4.5  \\
      & 1.20 &  8.0  \\
      & 1.20 & 10.0  \\
      & 0.70 & 10.0  \\
\hline
box 2 & 0.70 & 4.5  \\
      & 0.90 & 4.5  \\
      & 0.81 & 3.5  \\
      & 0.64 & 3.5  \\

\enddata
\end{deluxetable}

\clearpage
\begin{deluxetable}{lrrrrrrrrrrrrrrrrrrrrrl}
\tabletypesize{\tiny}
\rotate 
\setlength{\tabcolsep}{0.02in}

\tablecaption{Photometric distance estimates for new candidate nearby star
  systems\label{listnew}}

\tablewidth{0pt}
\tablehead{
\colhead{Name}                 &
\colhead{RA}                   &
\colhead{DEC}                  &
\colhead{pmra}                 &
\colhead{error}                &
\colhead{pmdc}                 &
\colhead{error}                &
\colhead{B}                    &
\colhead{V}                    &
\colhead{g}                    &
\colhead{r}                    &
\colhead{i}                    &
\colhead{J}                    &
\colhead{H}                    &
\colhead{K$_s$}                &
\colhead{W1}                   &
\colhead{W2}                   &
\colhead{W3}                   &
\colhead{W4}                   &
\colhead{Dist.est}             &
\colhead{error}                &
\colhead{colors}                &
\colhead{notes}                \\

\colhead{     }                &
\colhead{[deg]}                &
\colhead{[deg]}                &
\colhead{[mas/yr]}             &
\colhead{[mas/yr]}             &
\colhead{[mas/yr]}             &
\colhead{[mas/yr]}             &
\colhead{[mag]}                &
\colhead{[mag]}                &
\colhead{[mag]}                &
\colhead{[mag]}                &
\colhead{[mag]}                &
\colhead{[mag]}                &
\colhead{[mag]}                &
\colhead{[mag]}                &
\colhead{[mag]}                &
\colhead{[mag]}                &
\colhead{[mag]}                &
\colhead{[mag]}                &
\colhead{[pc]}                 &
\colhead{[pc]}                 &
\colhead{\#}                   &
\colhead{     }                }

\startdata

TYC 3980-1081-1                & 327.9096831  &   59.2941733 &   -79.9 &   0.8 &       69.1 &   0.8 &  11.958 &  10.501 &  11.185 &   9.980 &   8.600 &  6.529 &   5.860 &   5.651 &  5.320 &  5.150 &  5.254 &  5.140 &  5.93 &    1.32 &  16 & \tablenotemark{a}, \tablenotemark{b}, \tablenotemark{d}  \\
L 173-19                       &  30.1595833  &  -55.9679703 &   118.5 &   1.5 &      -69.1 &   1.1 &  13.467 &  11.877 &  12.617 &  11.270 &   9.809 &  7.625 &   7.088 &   6.773 &  6.619 &  6.382 &  6.362 &  6.176 &  8.47 &    1.50 &  16 & \tablenotemark{a}                                        \\
2MASS J20490993-4012062        & 312.2914328  &  -40.2017892 &   -51.4 &   1.5 &        1.7 &   1.5 &  15.294 &  13.532 &  14.278 &  12.825 &  10.994 &  8.596 &   8.018 &   7.704 &  7.474 &  7.295 &  7.163 &  7.123 &  8.66 &    1.48 &  16 & \tablenotemark{a}, \tablenotemark{b}                     \\
BPS CS 22898-0066              & 317.5192247  &  -19.3417936 &    87.0 &   1.4 &      -94.4 &   3.1 &  14.673 &  13.143 &  13.872 &  12.546 &  10.880 &  8.428 &   7.883 &   7.552 &  7.305 &  7.161 &  7.067 &  6.768 &  9.58 &    2.18 &  16 & \tablenotemark{a}, \tablenotemark{b}                     \\
TYC 3251-1875-1                &   3.9126392  &   47.5894489 &   -83.0 &   1.2 &      -20.7 &   1.2 &  12.605 &  11.120 &  11.804 &  10.511 &   9.234 &  7.336 &   6.709 &   6.467 &  6.349 &  6.107 &  6.132 &  6.003 &  9.66 &    1.81 &  16 & \tablenotemark{a}, \tablenotemark{b}                     \\
UPM 0815-2344                  & 123.7966400  &  -23.7376783 &   102.4 &   2.3 &       66.4 &   2.4 &  13.654 &  12.304 &  12.941 &  11.696 &  10.237 &  8.107 &   7.513 &   7.215 &  6.983 &  6.892 &  6.774 &  6.683 & 11.11 &    2.60 &  16 & \tablenotemark{a}                                        \\
2MASS J11473762+0501106        & 176.9067664  &    5.0196183 &    51.8 &   2.5 &       19.6 &   3.7 &  13.927 &  12.383 &  13.090 &  11.769 &  10.361 &  8.223 &   7.644 &   7.381 &  7.155 &  7.062 &  6.965 &  6.852 & 11.92 &    2.17 &  16 & \tablenotemark{a}, \tablenotemark{b}, \tablenotemark{d}  \\
UPM 1012-3124                  & 153.0378781  &  -31.4125789 &   -74.8 &   1.1 &       -9.4 &   1.0 &  15.052 &  13.460 &  14.165 &  12.879 &  11.189 &  8.848 &   8.262 &   7.993 &  7.758 &  7.497 &  6.128 &  4.798 & 12.08 &    2.41 &  16 & \tablenotemark{a}                                        \\
UPM 1951-3100                  & 297.9184097  &  -31.0060642 &    33.9 &   2.0 &     -148.0 &   0.9 &  13.929 &  12.307 &  13.056 &  11.654 &  10.148 &  8.251 &   7.698 &   7.409 &  7.224 &  7.069 &  6.975 &  6.845 & 12.18 &    2.76 &  16 & \tablenotemark{a}                                        \\
V* V549 Hya                    & 173.1715497  &  -26.8692036 &   -95.3 &   2.2 &      -28.6 &   4.7 &  17.235 &  15.647 &  16.425 &  14.800 &  13.080 &  9.837 &   9.276 &   9.012 &  8.928 &  8.645 &  8.481 &  8.536 & 12.36 &    4.37 &  16 & \tablenotemark{a}, \tablenotemark{b}                     \\
SCR 0757-7114                  & 119.3855169  &  -71.2482950 &    86.7 &   1.3 &       20.6 &   1.3 &  13.926 &  12.394 &  13.082 &  11.833 &  10.340 &  8.319 &   7.745 &   7.423 &  7.172 &  7.127 &  7.038 &  6.910 & 12.39 &    2.44 &  16 & \tablenotemark{a}, \tablenotemark{d}                     \\
EM* StHA 10                    &  28.2390044  &    8.5574914 &    96.0 &   4.4 &       15.5 &   4.5 &  15.641 &  14.055 &  14.754 &  13.448 &  11.575 &  9.238 &   8.638 &   8.358 &  8.216 &  8.009 &  7.864 &  7.739 & 12.57 &    2.52 &  16 & \tablenotemark{a}, \tablenotemark{b}, \tablenotemark{d}  \\
LHS 1918                       & 112.0545703  &  -18.7932367 &   -11.5 &   8.0 &      613.7 &   8.0 &  15.365 &  13.668 &  14.411 &  13.035 &  11.322 &  9.049 &   8.454 &   8.167 &  7.995 &  7.785 &  7.657 &  7.385 & 12.61 &    2.06 &  16 & \tablenotemark{a}                                        \\
PPMX 010320.9-134823           &  15.8373589  &  -13.8064025 &    76.5 &   6.1 &       43.9 &   6.0 &  13.544 &  12.026 &  12.748 &  11.441 &  10.152 &  8.138 &   7.571 &   7.258 &  7.032 &  7.002 &  6.929 &  6.791 & 13.15 &    2.37 &  16 & \tablenotemark{a}                                        \\
LTT 5790                       & 219.4733411  &  -34.6547450 &  -216.4 &   8.0 &     -103.7 &   8.0 &  14.423 &  12.920 &  13.593 &  12.335 &  10.835 &  8.674 &   8.003 &   7.762 &  7.611 &  7.485 &  7.366 &  7.138 & 13.26 &    2.79 &  16 & \tablenotemark{a}                                        \\
2MASS J01283952-1458042        &  22.1647592  &  -14.9679017 &   -17.1 &   3.8 &      -30.7 &   5.0 &  15.346 &  13.645 &  14.441 &  12.991 &  11.360 &  9.058 &   8.555 &   8.198 &  8.013 &  7.819 &  7.693 &  7.652 & 13.28 &    2.38 &  16 & \tablenotemark{a}, \tablenotemark{b}                     \\
UPM 1845-2855                  & 281.4895431  &  -28.9314258 &   -59.4 &   5.3 &      -93.1 &  20.2 &  14.067 &  12.611 &  13.231 &  12.044 &  10.538 &  8.414 &   7.897 &   7.603 &  7.410 &  7.265 &  7.190 &  7.030 & 13.29 &    2.87 &  16 & \tablenotemark{a}                                        \\
TYC 4823-2265-1                & 108.2967531  &   -5.1968422 &   -19.1 &   3.7 &     -304.7 &   4.8 &  12.730 &  11.183 &  11.955 &  10.533 &   9.496 &  7.649 &   7.077 &   6.819 &  6.707 &  6.478 &  6.464 &  6.320 & 13.44 &    2.52 &  15 & \tablenotemark{b}, \tablenotemark{c}, \tablenotemark{d}  \\
UPM 1816-5844                  & 274.0515567  &  -58.7348769 &    -4.2 &   2.2 &     -139.5 &   2.2 &  14.268 &  12.743 &  13.471 &  12.127 &  10.687 &  8.599 &   7.964 &   7.699 &  7.576 &  7.417 &  7.299 &  7.069 & 13.55 &    2.54 &  16 & \tablenotemark{a}, \tablenotemark{b}                     \\
UPM 0402-0242B                 &  60.6350025  &   -2.7093081 &    34.8 &   3.2 &      -38.2 &   2.0 &  14.022 &  12.458 &  13.180 &  11.885 &  10.516 &  8.451 &   7.830 &   7.552 &  7.387 &  7.240 &  7.167 &  7.006 & 13.72 &    2.52 &  16 & \tablenotemark{a}, \tablenotemark{d}                     \\
UPM 2303-4650                  & 345.8983450  &  -46.8464178 &  -171.6 &   1.1 &      -20.1 &   1.1 &  15.545 &  13.848 &  14.591 &  13.241 &  11.534 &  9.217 &   8.686 &   8.360 &  8.192 &  7.984 &  7.846 &  7.724 & 13.92 &    2.35 &  16 & \tablenotemark{a}                                        \\
G 162-70                       & 158.7546747  &   -9.4107175 &   222.7 &   8.7 &     -179.8 &  10.5 &  13.677 &  12.153 &  12.869 &  11.574 &  10.265 &  8.276 &   7.678 &   7.393 &  7.140 &  7.118 &  7.039 &  6.865 & 13.93 &    2.49 &  16 & \tablenotemark{a}                                        \\
UPM 1709-5957                  & 257.4430214  &  -59.9604250 &   -43.2 &   2.2 &      -60.0 &   2.2 &  14.975 &  13.411 &  14.069 &  12.790 &  11.131 &  9.003 &   8.388 &   8.127 &  7.943 &  7.759 &  7.629 &  7.507 & 14.13 &    2.77 &  16 & \tablenotemark{a}, \tablenotemark{b}                     \\
UPM 0838-2843                  & 129.6405922  &  -28.7239958 &    79.2 &   1.5 &     -129.2 &   1.5 &  13.508 &  11.864 &  12.642 &  11.239 &  10.031 &  8.108 &   7.584 &   7.282 &  7.000 &  6.931 &  6.854 &  6.704 & 14.22 &    2.75 &  16 & \tablenotemark{a}                                        \\
UPM 0409-4435                  &  62.3847642  &  -44.5943072 &   -20.9 &   1.3 &      113.2 &   1.3 &  15.923 &  14.250 &  14.993 &  13.647 &  11.946 &  9.464 &   8.912 &   8.561 &  8.381 &  8.191 &  8.032 &  7.915 & 14.31 &    3.00 &  16 & \tablenotemark{a}                                        \\
2MASS J06134717-2354250        &  93.4465839  &  -23.9069047 &   -33.9 &   2.4 &      109.1 &   1.0 &  14.521 &  12.952 &  13.679 &  12.359 &  10.926 &  8.723 &   8.160 &   7.872 &  7.693 &  7.538 &  7.437 &  7.386 & 14.34 &    2.74 &  16 & \tablenotemark{a}                                        \\
UPM 0415-4602                  &  63.9545317  &  -46.0399694 &   -93.3 &   5.3 &      -86.9 &   1.7 &  13.400 &  11.847 &  12.591 &  11.271 &   9.998 &  8.120 &   7.542 &   7.293 &  7.074 &  6.988 &  6.896 &  6.766 & 14.46 &    2.69 &  16 & \tablenotemark{a}                                        \\
LP 489-3                       & 277.3510525  &  -34.9625439 &   -52.4 &   8.0 &     -236.0 &   8.0 &  13.353 &  11.877 &  12.578 &  11.286 &   9.997 &  8.086 &   7.565 &   7.327 &  7.146 &  6.978 &  6.930 &  6.846 & 14.75 &    2.93 &  16 & \tablenotemark{a}                                        \\

\enddata

\tablecomments{Only a portion of this table is shown here to demonstrate its form and content. A machine-readable version of the full table is available in the electronic edition.}
\tablenotetext{a} {candidate selected from box 1, using Figure 3}
\tablenotetext{b} {candidate within 30 arcseconds of an X-ray source}
\tablenotetext{c} {candidate selected from box 2, using Figure 3}
\tablenotetext{d} {candidate's motion could not be verified by eye}

\end{deluxetable}

\clearpage
\begin{deluxetable}{lrrrrrrrrrrrrrrrrrrrrrrcl}
\tabletypesize{\tiny}
\rotate 
\setlength{\tabcolsep}{0.02in}

\tablecaption{Photometric distance estimates for systems recovered in
  the nearby star search\label{recover}}

\tablewidth{0pt}
\tablehead{
\colhead{Name}                 &
\colhead{RA}                   &
\colhead{DEC}                  &
\colhead{pmra}                 &
\colhead{error}                &
\colhead{pmdc}                 &
\colhead{error}                &
\colhead{B}                    &
\colhead{V}                    &
\colhead{g}                    &
\colhead{r}                    &
\colhead{i}                    &
\colhead{J}                    &
\colhead{H}                    &
\colhead{K$_s$}                &
\colhead{W1}                   &
\colhead{W2}                   &
\colhead{W3}                   &
\colhead{W4}                   &
\colhead{Dist.est}             &
\colhead{Error}                &
\colhead{Colors}               &
\colhead{Dist.pub}             &
\colhead{Dist.type}            &
\colhead{Notes}                \\

\colhead{     }                &
\colhead{[deg]}                &
\colhead{[deg]}                &
\colhead{[mas/yr]}             &
\colhead{[mas/yr]}             &
\colhead{[mas/yr]}             &
\colhead{[mas/yr]}             &
\colhead{[mag]}                &
\colhead{[mag]}                &
\colhead{[mag]}                &
\colhead{[mag]}                &
\colhead{[mag]}                &
\colhead{[mag]}                &
\colhead{[mag]}                &
\colhead{[mag]}                &
\colhead{[mag]}                &
\colhead{[mag]}                &
\colhead{[mag]}                &
\colhead{[mag]}                &
\colhead{[pc]}                 &
\colhead{[pc]}                 &
\colhead{\#}                   &
\colhead{[pc]}                 &
\colhead{     }                &
\colhead{     }                }

\startdata

PM I00026+3821          &  0.6671503   &   38.3625119 &    -76.2 &   3.9 &    -25.3 &  3.6 &  15.507 &  13.931 & 14.671 & 13.303 & 11.785 & 9.707 &  9.203 &  8.911 & 8.568 &  8.385 &  8.265 &  8.506 &  22.98 &    4.45 &    16 & 22.08  &  P  & \tablenotemark{a}                                       \\      
LP 644-34               &  1.3952731   &   -6.1186375 &    177.5 &   8.0 &    -64.4 &  8.0 &  14.732 &  13.147 & 13.923 & 12.526 & 11.275 & 9.255 &  8.649 &  8.411 & 8.219 &  8.094 &  7.983 &  7.856 &  22.02 &    3.91 &    16 & 26.11  &  P  & \tablenotemark{a}                                       \\
LHS 1019                &  1.5797275   &  -65.8403125 &    195.8 &   8.0 &   -552.5 &  8.0 &  13.731 &  12.188 & 12.917 & 11.574 & 10.367 & 8.479 &  7.839 &  7.631 & 7.416 &  7.336 &  7.226 &  6.995 &  17.14 &    3.06 &    16 & 16.71  &  T  & \tablenotemark{a}                                       \\
2MASS J00080642+4757025 &  2.0267508   &   47.9506631 &   -126.9 &   4.4 &      2.5 &  2.5 &  14.390 &  12.779 & 13.472 & 12.150 & 10.640 & 8.523 &  8.000 &  7.677 & 7.473 &  7.336 &  7.221 &  6.962 &  12.76 &    2.26 &    16 & 13.85  &  P  & \tablenotemark{a}, \tablenotemark{c}                    \\
BPM 46052               &  2.3331467   &  -21.2448286 &    138.5 &   1.2 &   -101.9 &  2.6 &  13.698 &  12.179 & 12.935 & 11.540 & 10.512 & 8.763 &  8.156 &  7.937 & 7.751 &  7.686 &  7.586 &  7.532 &  23.93 &    4.61 &    14 & 26.67  &  P  & \tablenotemark{b}                                       \\
GSC 04018-02763         &  2.6066942   &   62.2104169 &    -27.5 &   2.0 &     35.2 &  2.0 &  15.394 &  13.803 & 14.514 & 13.221 & 11.841 & 9.655 &  9.092 &  8.811 & 8.605 &  8.464 &  8.273 &  7.660 &  23.25 &    4.52 &    16 & 20.58  &  P  & \tablenotemark{a}                                       \\
BPS CS 30324-0025       &  2.6793694   &  -20.6519094 &    120.3 &   1.4 &    -75.5 &  1.3 &  14.909 &  13.392 & 14.115 & 12.757 & 11.467 & 9.480 &  8.860 &  8.614 & 8.491 &  8.359 &  8.251 &  8.288 &  24.05 &    4.25 &    16 & 23.81  &  P  & \tablenotemark{a}, \tablenotemark{c}                    \\
GJ 0011                 &  3.3163239   &   69.3270800 &    717.0 &   2.5 &   -292.4 &  2.5 &  13.363 &  12.487 & 13.198 & 11.827 & 10.520 & 8.556 &  7.984 &  7.746 & 7.562 &  7.411 &  7.318 &  7.208 &  18.33 &    6.83 &    16 & 21.05  &  T  & \tablenotemark{a}                                       \\
HIP 1083                &  3.3712489   &  -36.8286303 &   -213.8 &   8.0 &   -335.2 &  8.0 &  12.240 &  10.781 & 11.530 & 10.162 &  9.394 & 7.803 &  7.183 &  6.962 & 6.832 &  6.763 &  6.716 &  6.644 &  20.60 &    3.82 &     7 & 27.22  &  T  & \tablenotemark{b}                                       \\
LTT 17095               &  3.4112875   &   80.6658033 &    249.9 &   0.9 &    190.8 &  0.9 &  12.626 &  11.104 & 11.842 & 10.517 &  9.519 & 7.756 &  7.131 &  6.904 & 6.651 &  6.707 &  6.660 &  6.579 &  15.58 &    2.78 &    13 & 19.59  &  T  & \tablenotemark{b}, \tablenotemark{c}                    \\
G 242-049               &  3.9031153   &   72.2837533 &    319.0 &   8.0 &    185.0 &  8.0 &  13.727 &  12.362 & 12.933 & 11.707 & 10.593 & 8.837 &  8.250 &  7.991 & 7.859 &  7.760 &  7.682 &  7.626 &  24.29 &    5.26 &    15 & 19.80  &  P  & \tablenotemark{b}                                       \\
LHS 1049                &  3.9305203   &  -67.9932772 &    611.2 &   8.0 &   -152.9 &  8.0 &  14.064 &  12.531 & 13.286 & 11.961 & 10.752 & 8.804 &  8.257 &  8.020 & 7.824 &  7.694 &  7.598 &  7.564 &  20.81 &    3.59 &    16 & 20.12  &  P  & \tablenotemark{a}                                       \\
GJ 0012                 &  3.9553561   &   13.5562761 &    621.0 &   8.0 &    333.0 &  8.0 &  14.265 &  12.600 & 13.405 & 11.963 & 10.698 & 8.619 &  8.068 &  7.807 & \nodata & \nodata & \nodata & \nodata &  15.87 &    3.03 &    16 & 11.57  &  T  & \tablenotemark{a}                                       \\
LHS 1051                &  3.9640403   &  -67.9976672 &    611.2 &   8.0 &   -152.9 &  8.0 &  12.375 &  10.915 & 11.664 & 10.562 &  9.458 & 7.795 &  7.211 &  6.949 & 6.840 &  6.747 &  6.704 &  6.572 &  18.18 &    4.39 &    12 & 21.46  &  P  & \tablenotemark{b}                                       \\
2MASS J00155808-1636578 &  3.9919686   &  -16.6160100 &   -110.2 &   2.2 &     35.0 &  2.2 &  14.875 &  13.184 & 13.976 & 12.547 & 11.027 & 8.736 &  8.191 &  7.909 & 7.690 &  7.546 &  7.435 &  7.214 &  12.66 &    2.33 &    16 & 17.95  &  T  & \tablenotemark{a}, \tablenotemark{c}, \tablenotemark{d} \\

\enddata

\tablecomments{Only a portion of this table is shown here to demonstrate its form and content. A machine-readable version of the full table is available in the electronic edition.}
\tablenotetext{a} {candidate selected from box 1, using Figure 3}
\tablenotetext{b} {candidate selected from box 2, using Figure 3}
\tablenotetext{c} {candidate within 30 arcseconds of an X-ray source}
\tablenotetext{d} {distance estimate from this survey differs by more than 2 sigma from published parallax or distance estimate} 
\tablenotetext{e} {candidate is a known giant}

\end{deluxetable}

\clearpage




\begin{thebibliography}{}

\bibitem[Bergeat et al.(2002)]{Bergeat} Bergeat, J., Knapik, A., \&
  Rutily, B.\ 2002, VizieR Online Data Catalog, 339, 967

\bibitem[Bochanski et al.(2005)]{Bochanski} Bochanski, J.~J., Hawley,
  S.~L., Reid, I.~N., et al.\ 2005, \aj, 130, 1871

\bibitem[Boyd et al.(2011a)]{TSNXXVII} Boyd, M.~R., Henry, T.~J., Jao,
  W.-C., Subasavage, J.~P., \& Hambly, N.~C.\ 2011a, \aj, 142, 92

\bibitem[Boyd et al.(2011b)]{TSNXXV} Boyd, M.~R., Winters, J.~G.,
  Henry, T.~J., et al.\ 2011b, \aj, 142, 10

\bibitem[Christlieb et al.(2001)]{Christlieb} Christlieb, N., Green,
  P.~J., Wisotzki, L., \& Reimers, D.\ 2001, VizieR Online Data
  Catalog, 337, 50366

\bibitem[Deacon et al.(2005)]{Deacon1} Deacon, N.~R.,
  Hambly, N.~C., \& Cooke, J.~A.\ 2005, \aap, 435, 363

\bibitem[Deacon \& Hambly(2007)]{Deacon2} Deacon, N.~R.,
  \& Hambly, N.~C.\ 2007, \aap, 468, 163

\bibitem[Dieterich et al.(2014)]{Dieterich} Dieterich,
  S.~B., Henry, T.~J., Jao, W.-C., et al.\ 2014, \aj, 147, 94

\bibitem[Dittmann et al.(2014)]{Dittmann} Dittmann, J.~A., Irwin,
  J.~M., Charbonneau, D., \& Berta-Thompson, Z.~K.\ 2014,
  arXiv:1312.3241

\bibitem[Finch et al.(2007)]{TSNXVIII} Finch, C.~T., Henry, T.~J.,
  Subasavage, J.~P., Jao, W.-C., \& Hambly, N.~C.\ 2007, \aj, 133,
  2898

\bibitem[Finch et al.(2010)]{upm1} Finch, C.~T., Zacharias, N., \&
  Henry, T.~J.\ 2010, \aj, 140, 844

\bibitem[Finch et al.(2012b)]{upm2} Finch, C.~T., Zacharias, N., Boyd,
  M.~R., Henry, T.~J., \& Hambly, N.~C.\ 2012b, \apj, 745, 118

\bibitem[Finch et al.(2012a)]{urataas} Finch, C.~T., Bredthauer, G.,
  DiVittorio, M., et al.\ 2012a, American Astronomical Society Meeting
  Abstracts \#220, 220, \#135.05

\bibitem[Gizis et al.(2002)]{Gizis} Gizis, J.~E., Reid, I.~N., \&
  Hawley, S.~L.\ 2002, \aj, 123, 3356

\bibitem[Gray et al.(2003)]{Gray1} Gray, R.~O., Corbally, C.~J.,
  Garrison, R.~F., McFadden, M.~T., \& Robinson, P.~E.\ 2003, \aj,
  126, 2048

\bibitem[Gray et al.(2006)]{Gray2} Gray, R.~O., Corbally, C.~J.,
  Garrison, R.~F., et al.\ 2006, \aj, 132, 161

\bibitem[Hambly et al.(2004)]{TSNVIII} Hambly, N.~C., Henry, T.~J.,
  Subasavage, J.~P., Brown, M.~A., \& Jao, W.-C.\ 2004, \aj, 128, 437

\bibitem[Hawley et al.(2002)]{Hawley2} Hawley, S.~L., Covey, K.~R.,
  Knapp, G.~R., et al.\ 2002, \aj, 123, 3409

\bibitem[Henry et al.(2006)]{TSNXVII} Henry, T.~J., Jao, W.-C.,
  Subasavage, J.~P., et al.\ 2006, \aj, 132, 2360

\bibitem[Henry et al.(2004)]{TSNX} Henry, T.~J., Subasavage, J.~P.,
  Brown, M.~A., et al.\ 2004, \aj, 128, 2460

\bibitem[Hoeg et al.(2000)]{Hoeg} Hoeg, E., Fabricius, C., Makarov,
  V.~V., et al.\ 2000, VizieR Online Data Catalog, 1259, 0

\bibitem[Jao et al.(2005)]{2005AJ....129.1954J} Jao, W.-C., Henry,
  T.~J., Subasavage, J.~P., et al.\ 2005, \aj, 129, 1954

\bibitem[Kastner et al.(1993)]{Kastner} Kastner, J.~H., Forveille, T.,
  Zuckerman, B., \& Omont, A.\ 1993, VizieR Online Data Catalog, 327,
  50163

\bibitem[Le Bertre et al.(2001)]{Bertre1} Le Bertre, T., Matsuura, M.,
  Winters, J.~M., et al.\ 2001, \aap, 376, 997

\bibitem[Le Bertre et al.(2003)]{Bertre2} Le Bertre, T., Tanaka, M.,
  Yamamura, I., \& Murakami, H.\ 2003, \aap, 403, 943

\bibitem[L{\'e}pine(2005)]{Lepine3} L{\'e}pine, S.\ 2005, \aj, 130,
  1680


\bibitem[L{\'e}pine \& Shara(2005)]{Lepine0} L{\'e}pine, S., \& Shara,
  M.~M.\ 2005, \aj, 129, 1483

\bibitem[L{\'e}pine(2005)]{Lepine2} L{\'e}pine, S.\ 2005,
  \aj, 130, 1247

\bibitem[L{\'e}pine(2008)]{Lepine5} L{\'e}pine, S.\ 2008, \aj, 
135, 2177

\bibitem[L{\'e}pine \& Gaidos(2011)]{Lepine1} L{\'e}pine, S., \&
  Gaidos, E.\ 2011, \aj, 142, 138

\bibitem[Luyten(1980)]{1980PMMin..55....1L} Luyten, W.~J.\ 1980,
  Proper Motion Survey with the 48-inch Telescope, Univ.~Minnesota,
  55, 1 (1980), 55, 1

\bibitem[Masseron et al.(2010)]{Masseron} Masseron, T., Johnson,
  J.~A., Plez, B., et al.\ 2010, VizieR Online Data Catalog, 350,
  99093

\bibitem[Perryman \& ESA(1997)]{Perryman} Perryman, M.~A.~C., \& ESA
  1997, ESA Special Publication, 1200,

\bibitem[Reid et al.(1997)]{Hawley1} Reid, I.~N., Hawley, S.~L., \&
  Gizis, J.~E.\ 1997, VizieR Online Data Catalog, 3198, 0

\bibitem[Reid et al.(2003)]{Reid1} Reid, I.~N., Cruz,
  K.~L., Allen, P., et al.\ 2003, \aj, 126, 3007

\bibitem[Reid et al.(2004)]{Reid2} Reid, I.~N., Cruz, K.~L., Allen,
  P., et al.\ 2004, \aj, 128, 463

\bibitem[Riaz et al.(2006)]{Riaz1} Riaz, B., Gizis, J.~E., \& Harvin,
  J.\ 2006, \aj, 132, 866

\bibitem[Riedel et al.(2010)]{TSNXXII} Riedel, A.~R., Subasavage,
  J.~P., Finch, C.~T., et al.\ 2010, \aj, 140, 897

\bibitem[Riedel(2012)]{Adric} Riedel, A.~R., Ph.D. thesis,
  Georgia\ 2012 State Univ.

\bibitem[Riedel et al.(2014)]{2014AJ....147...85R} Riedel, A.~R.,
  Finch, C.~T., Henry, T.~J., et al.\ 2014, \aj, 147, 85

\bibitem[Scholz et al.(2000)]{2000A&A...353..958S} Scholz, R.-D.,
  Irwin, M., Ibata, R., Jahrei{\ss}, H., \& Malkov, O.~Y.\ 2000, \aap,
  353, 958

\bibitem[Scholz et al.(2002)]{2002ApJ...565..539S} Scholz, R.-D.,
  Szokoly, G.~P., Andersen, M., Ibata, R., \& Irwin, M.~J.\ 2002,
  \apj, 565, 539

\bibitem[Subasavage et al.(2005a)]{TSNXII} Subasavage, J.~P., Henry,
  T.~J., Hambly, N.~C., Brown, M.~A., \& Jao, W.-C.\ 2005a, \aj, 129,
  413

\bibitem[Subasavage et al.(2005b)]{TSNXV} Subasavage, J.~P., Henry,
  T.~J., Hambly, N.~C., et al.\ 2005b, \aj, 130, 1658

\bibitem[Whitelock et al.(2006)]{Whitelock} Whitelock, P.~A., Feast,
  M.~W., Marang, F., \& Groenewegen, M.~A.~T.\ 2006, VizieR Online
  Data Catalog, 736, 90751

\bibitem[Zacharias et al.(2004)]{UCAC2} Zacharias, N., Urban, S.~E.,
  Zacharias, M.~I., et al.\ 2004, \aj, 127, 3043

\bibitem[Zacharias(2005)]{URAT} Zacharias, N.\ 2005, Astrometry in the
  Age of the Next Generation of Large Telescopes, 338, 98

\bibitem[Zacharias et al.(2010)]{UCAC3} Zacharias, N., Finch, C.,
  Girard, T., et al.\ 2010, \aj, 139, 2184

\bibitem[Zacharias et al.(2013)]{UCAC4} Zacharias, N., Finch, C.~T.,
  Girard, T.~M., et al.\ 2013, \aj, 145, 44

\end{thebibliography}
\end{document}